\documentclass[%
 superscriptaddress,
 preprint,
 amsmath,amssymb,
 aps,
]{revtex4-2}

\usepackage{graphicx}
\usepackage{float}
\usepackage{dcolumn}
\usepackage{bm}
\usepackage{empheq}
\usepackage{xcolor}
\usepackage{longtable}

\makeatletter
\def\active@comma{,\penalty\z@\relax}%
\def\fm@present@note#1#2{%
 \begingroup
  \csname c@\@mpfn\endcsname#1\relax
  \def\@thefnmark{\frontmatter@thefootnote}%
  \set@footnotefont
  \set@footnotewidth
  \@parboxrestore
  \def\baselinestretch{1}\selectfont
  \parskip\z@skip
  \parindent\z@ \leftskip\z@ \rightskip\z@
  \centering
  \@makefnmark\,\ignorespaces#2%
  \par
 \endgroup
}%
\def\fm@produce@notes{%
 \par
 \addvspace{6\p@}%
 \begingroup
  \let\@TBN@opr\fm@present@note
  \@FMN@list
 \endgroup
 \global\let\@FMN@list\@empty
}%
\@ifundefined{preprintsty@sw}{}{%
 \@ifundefined{frontmatter@footnote@produce}{}{%
  \preprintsty@sw{\let\frontmatter@footnote@produce\fm@produce@notes}{}%
 }%
}%
\makeatother

\begin{document}

\preprint{}

\title{An Inverse Grad-Shafranov Neural Network Approach to Tokamak Magnetic Control}

\newcommand{\PSFC}{MIT Plasma Science and Fusion Center, Cambridge, Massachusetts 02139, USA}
\newcommand{\SPC}{Swiss Plasma Center, École Polytechnique Fédérale de Lausanne (EPFL),\\ CH-1015 Lausanne, Switzerland}

\author{Allen M. Wang}\email{awang@psfc.mit.edu}
\affiliation{\PSFC}
\author{Adriano Mele}\email{adriano.mele@epfl.ch}%
\affiliation{\SPC}
\author{Cosmas Hei\ss{}}
\affiliation{\SPC}
\author{Cristian Galperti}
\affiliation{\SPC}
\author{Zander Keith}
\affiliation{\PSFC}
\author{Alessandro Pau}
\affiliation{\SPC}
\author{Antoine Merle}
\affiliation{\SPC}
\author{Olivier Sauter}
\affiliation{\SPC}
\author{Daniel Gonzalez Casti\~neiras}
\affiliation{\SPC}
\author{Francesco Carpanese}
\affiliation{\SPC}
\author{Federico Felici}
\altaffiliation{Work done while at Google DeepMind, London, UK.}
\affiliation{Fusionality SA, Lausanne, Switzerland}
\author{Mark Dan Boyer}
\affiliation{Commonwealth Fusion Systems, Devens, Massachusetts 01434, USA}
\author{Cristina Rea}
\affiliation{\PSFC}
\author{the TCV Team}
\affiliation{See author list of C. Theiler \textit{et al.}~\cite{theiler2026progress}}
\author{the EUROfusion Tokamak Exploitation Team}
\affiliation{See author list of N. Vianello \textit{et al.}~\cite{vianello2026results}}

\date{\today}

\begin{abstract}
A new approach to tokamak magnetic control enabling high-precision plasma shaping and novel real-time adaptability is experimentally demonstrated on the Tokamak à Configuration Variable (TCV). The method is motivated by the insight that, under appropriate assumptions, a real-time inverse Grad–Shafranov solver approximates an optimal control policy for plasma boundary regulation. Building on this, a control architecture is developed in which classical controllers enforce operational constraints while a fast surrogate model provides a real-time inverse mapping from the desired plasma boundary to Poloidal Field Coil currents.

Experimental results on TCV demonstrate improved plasma shaping with respect to the standard discharge preparation procedure --- albeit without explicit real-time shape feedback --- while enabling flexible response to asynchronous events. It is shown that a single network provides satisfactory performance across a range of plasma magnetic configurations. Real-time adaptivity is demonstrated in simulation, and partially in experiment, through adaptive strike point motion and early termination in response to a real-time trigger.

These results suggest a viable path toward magnetic control architectures that reduce reliance on dense diagnostic coverage while maintaining high-accuracy plasma shaping, with potential relevance for future fusion power plant operation.
\end{abstract}

\maketitle
\section{Introduction}
Magnetic control is a relatively mature subject with respect to the requirements of contemporary devices ~\cite{hofmann1994creation}, yet several open challenges remain. Upcoming breakeven tokamaks will likely demand unprecedented control precision and robustness. Moreover, real-time adaptation of plasma shape and current trajectories may become necessary for managing heat flux~\cite{frattolillo2025magnetic} and off-normal events~\cite{wang2025learning} within prescribed safety limits~\cite{frattolillo2025implementation}. Conventional approaches may also prove difficult or infeasible to deploy in a reactor environment, as they typically rely on in-vessel diagnostics unlikely to survive the high neutron fluence expected in fusion power plants~\cite{gonccalves2023advances}. Together, these challenges motivate continued advances in magnetic control.

In many tokamak experiments, plasma shaping is primarily obtained by designing suitable coil currents through the solution of the Inverse Grad-Shafranov (IGS) problem.
Given a description of the desired magnetic geometry, described for instance through a collection of boundary and X points, along with a set of internal plasma state variables (e.g. total current, poloidal beta, internal inductance), the solution to an IGS consists in finding a plasma equilibrium that matches the desired geometry and internal parameters as closely as possible. This is typically done by assuming a given profile of the internal plasma toroidal current distribution across the nested magnetic surfaces inside the separatrix. One of the main limitations of this approach is that the exact internal current profile, as well as some commonly used global internal parameters related to it, such as the poloidal beta and the internal inductance, are usually unknown \emph{a priori}. This mismatch can ultimately lead to a non-negligible difference between the pre-programmed and obtained plasma shape.

To mitigate this issue, one possibility is to adopt closed-loop plasma shape control. Classic techniques involve Multi-Input, Multi-Output (MIMO) PID-based solutions~\cite{ariola2005plasma,mele2019mimo,mele2025design}. Moreover, the use of optimization-based techniques is also emerging as a potential path to advance tokamak magnetic control. One approach is model predictive control (MPC)~\cite{gerkvsivc2018model,tartaglione2022plasma}, which has recently been demonstrated experimentally in TCV~\cite{mele2025first}. Another is end-to-end deep Reinforcement Learning (RL), which was also first demonstrated in TCV and has garnered considerable attention~\cite{degrave2022magnetic,tracey2024towards}. While both approaches show promise, each of them also presents specific limitations. For example, the prediction model included in the online optimization loop of the MPC approach is based on a linearization of the plasma dynamics close to the target configuration, making the inclusion of nonlinear effects challenging. RL-based approaches, on the other hand, raise concerns regarding transfer from simulation to reality and compatibility with machine protection considerations.

This paper analyzes the magnetic control problem from an optimization perspective, formulating a Markov Decision Process (MDP) corresponding to the control problem. Analysis of the MDP motivates a control policy combining a real-time IGS solver with a classical RZIP feedback controller. To enable real-time execution, a neural network surrogate of the IGS solver is employed. This neural network surrogate takes a description of the desired plasma shape, along with real-time estimates of parameters related to the plasma internal current distribution, and outputs optimized coil currents. These coil currents are sent to the RZIP controller as control targets, providing a layer of separation between the neural network and basic machine control functions and hardware that allows for machine protection considerations. Since the target shape is an input to the real-time neural network, significant real-time adaptation of the plasma shape is possible, enabling novel capabilities for adapting the plasma magnetic configuration in response to real-time events.

The resulting control policy is demonstrated in both experiment and simulation. Precise control of a Lower Single Null (LSN) plasma with rejection of the disturbance related to variations in $\beta_p$ is shown experimentally. Double Null (DN) plasmas are also controlled experimentally, with favorable control precision when compared to conventional linear control approaches. Real-time adaptation capabilities are demonstrated in simulation, with partial demonstration in experiment. Simulations further show that a single neural network can control a diversity of plasma shapes, spanning LSN, Upper Single Null (USN), DN, and both Positive and Negative Triangularity configurations.



\clearpage
\section{Theoretical formulation and control design}
This section presents a theoretical analysis motivating the control strategy developed in this work. Sec.~\ref{subsec:modelling_assumptions} begins by stating the physics modelling assumptions. Sec.~\ref{subsec:time_dep_prob} defines the time-dependent control optimization problem to be solved. Sec.~\ref{subsec:time_indep_relaxation} shows that, under a set of assumptions, the time-dependent optimal control problem can be approximated by a sequence of time-independent optimization problems solved in real-time at every time step. Sec.~\ref{subsec:practical_controller} discusses practical approaches to relaxing the assumptions in~\ref{subsec:time_indep_relaxation}, leading to the control strategy demonstrated experimentally; limitations of the assumptions underlying the analysis are also discussed. Finally, Sec.~\ref{subsec:rl_connection} connects the proposed approach theoretically to the RL-based method in~\cite{degrave2022magnetic,tracey2024towards}.

See Table~\ref{tab:symb} for a summary of the notation used in this section.

\subsection{Modelling Assumptions}\label{subsec:modelling_assumptions}
This work adopts a set of modelling assumptions commonly used for tokamak magnetic control. Namely, we consider a model that couples the Grad-Shafranov equation \cite{Grad1958,shafranov1958magnetohydrodynamical} to circuit equations that govern the time evolution of external conductors. The Grad-Shafranov equation, in COCOS=17 convention \cite{sauter2013tokamak}, is given by:
\begin{align}\label{eq:GS}
    &\Delta^* \psi = -2\pi R\mu_0 j_\phi = -4\pi^2\left(\mu_0R^2 p'(\psi) + TT'(\psi)\right) \quad &\text{in the plasma cross section}\,, \\
    &\Delta^* \psi = 0 \quad &\text{in vacuum} \,, \\
    &\Delta^* \psi = j_e \quad &\text{in the external conductors} \,, 
\end{align}
where $j_\phi$ and $j_e$ are the toroidal current densities in the plasma and the external conductors respectively, $p(\psi)$ is the pressure profile as a function of the poloidal flux $\psi$, and the function $T(\psi) = rB_t$, related to the diamagnetic effect of the plasma, is sometimes called the poloidal current function. In this work, we let $\bm{\theta}$ denote the plasma parameters used to represent the pressure and current distributions and treat this as an exogenous vector. 
The current density in the external conductors is related to the total currents $\mathbf{I}_e$, which are further partitioned into shaping coil directions, $\mathbf{I}_s$, Ohmic drive coil directions, $\mathbf{I}_o$, and passive conductor currents, $\mathbf{I}_v$:
\begin{align}
    \mathbf{I}_e = [\mathbf{I}_s, \mathbf{I}_o, \mathbf{I}_v] = [\mathbf{I}_a, \mathbf{I}_v] \,.
\end{align}
The shaping coils and the transformer stack together form the active coils, whose currents are denoted by $\mathbf{I}_a$. Note that, in this formulation, we are neglecting any in-vessel coil that is exclusively used for vertical stabilization. We will assume that the plasma is already stabilized, and that any shaping action is performed on slower time-scales than those typical of the vertical stabilization system to avoid undesired interactions.

The time evolution of external conductors is governed by the circuit equation:
\begin{align}\label{eq:circuit_dic}
    M_{ee}\dot{\mathbf{I}}_e +  R_e\mathbf{I}_e + \dot{\Psi}_{ep} = \begin{bmatrix}
        \mathbf{V}_a\\0
    \end{bmatrix}
\end{align}
where $M_{ee}$ is the mutual inductance matrix of the conductors, $\dot{\Psi}_{ep}$ is the induced emf due to coupling with the plasma, $R_e$ is the diagonal resistance matrix, and $\mathbf{V}_a$ are the voltage actions applied to active coils.

Given a vector of external conductor currents $\mathbf{I}_e$ and plasma parameters $\bm{\theta}$, the Grad-Shafranov equation can be numerically solved to yield a poloidal flux distribution $\psi$:
\begin{align}
    \psi = F(\mathbf{I}_e, \bm{\theta})
\end{align}
It should be noted that the GS solution is not unique in general, and recent literature highlights the need for a better understanding of the consequences~\cite{Pentland_2025}. Examination of GS solution uniqueness is outside the scope of this paper; in practical scenarios, it is standard to assume it.

\subsection{The Time-Dependent Problem}\label{subsec:time_dep_prob}
We consider the control objective of achieving a target plasma shape, defined by a vector of shape descriptors $\mathbf{s}$. The shape of a fusion plasma is usually defined as the Last Closed Flux Surface (LCFS) in the vacuum chamber poloidal cross-section; common shape descriptors often used in control include geometric momenta of the LCFS such as elongation, triangularity and squareness, geometric quantities such as plasma–wall gaps, or magnetic fluxes and field components evaluated at selected control locations. We assume that these descriptors are computed using some shape observer $S$, such that $\mathbf{s} = S(\psi)$. 
The control objective is to minimize the error with respect to user-specified targets $\mathbf{s}^{\text{targ}}$ according to a distance metric $d(\mathbf{s}, \mathbf{s}^{\text{targ}})$.
We now state the general time-dependent optimization problem over a $N$-step time horizon. This problem involves finding actions, in terms of voltages applied to the active coils, that minimize the error with respect to a trajectory of shape targets, while satisfying the constraints of our physical model. 
These voltages can be obtained through a suitable control policy $\pi$ that maps state to actions.
Letting the subscript $t$ denote the current time step, the general time-dependent problem is given by:
\begin{subequations}\label{eq:mdp}
    \begin{align}
        \min_{\pi} &\sum_{t=1}^N d(\mathbf{s}_t, \mathbf{s}^{\text{targ}}_{t})\\
        s.t. \quad&\mathbf{s}_t = S(F(\mathbf{I}_{e, t},\bm{\theta}_t))\\
        & \mathbf{I}_{e,t+1} \text{ per Eq.~\eqref{eq:circuit_dic}}\\
        & { \mathbf{V}_{a,t} = \pi(\mathbf{s}_t, \mathbf{s}^{\text{targ}}_{t},\mathbf{I}_{e, t}, \dot{\psi}_{ep,t}, \bm{\theta}_t) }
    \end{align}
\end{subequations}
In this formulation, the actions are voltages applied to the active coils at each time step, $\mathbf{V}_{a,t}$. These in turn affect the currents $\mathbf{I}_{e,t+1}$ through Eq.~\eqref{eq:circuit_dic}. This was the choice adopted for the RL policy in~\cite{degrave2022magnetic}. {Note that, in this formulation, the policy $\pi$ will depend in general on the measured and target plasma shapes, on the currents in the external conductors, on the voltage induced by the plasma on the coils and on the estimated plasma state.} 
This is a nonlinear time-dependent control optimization problem which is computationally challenging to solve in general. Note that the superficial change of converting $\min d(\cdot, \cdot)$ into $\max -d(\cdot, \cdot)$ makes Eq.~\eqref{eq:mdp} an MDP, which can be addressed with methods from RL. This connection is expanded upon in Sec.~\ref{subsec:rl_connection}.

The aim of the next section is to relax this formulation so as to only retain the static part of the IGS problem; this way, the actions of the required policy are the currents in the shaping circuits instead.


\subsection{A Time-Independent Relaxation}\label{subsec:time_indep_relaxation}
The time-dependence in Eq.~\eqref{eq:mdp} is encapsulated in the circuit equations~\eqref{eq:circuit_dic}. Under the assumption that the coil currents in Eq.~\eqref{eq:circuit_dic} are perfectly controlled by a low-level coil current controller and that vessel currents can be neglected on the timescales of interest, the circuit equations can be removed from the optimization problem. This results in a significant simplification of the optimization problem, now time-independent, the optimized variables being the active coil currents $\mathbf{I}_{a}$. 
%
Further restricting Ohmic coils from being used for shape target optimization, the resulting minimization objective is simplified to finding the optimal $\mathbf{I}_s$:
\begin{align}
    &\sum_{t=1}^N \min_{\mathbf{I}_{s,t}} d(\mathbf{s}_t, \mathbf{s}^{\text{targ}}_{t})\\
    s.t. \quad&\mathbf{s}_t = S(F(\mathbf{I}_{s,t},\mathbf{I}_{o,t},\mathbf{I}_{v,t},\bm{\theta}_t)).
\end{align}
In practice, tokamak coil currents are usually controlled to a reasonably good degree of accuracy and on time scales significantly faster than the typical ones adopted for shape control. In fact, poloidal field coil currents are often considered as the main actuators for plasma shape controllers, often designed based on the static steady-state relationship between active shaping currents $\mathbf{I}_s$ and shape descriptors $\mathbf{s}$ (as, for example, in the case of ITER's proposed Current Control Scheme~\cite{mattei2026recent}). Hence, this assumption constitutes a reasonable approximation in many cases of practical relevance. 

{Moreover, especially when the plasma shape is regulated to a steady-state configuration such as during a typical discharge flat-top, vessel currents are often neglected when designing shape control algorithms. However, their role might still be relevant during more dynamic phases, such as controlled ramp-downs. In this perspective}, it is worth observing that the shaping coils can be used to directly counteract the poloidal fields introduced by $\mathbf{I}_o$ and $\mathbf{I}_v$. Letting $B_{*}$ be the matrix mapping currents in a given set of coils $*$ to the corresponding poloidal magnetic field, the poloidal stray fields due to $\mathbf{I}_o$ and $\mathbf{I}_v$ are given by:
\begin{align}
    \mathbf{B}_{\text{stray}} = B_o \mathbf{I}_o + B_v \mathbf{I}_v \,.
\end{align}
Perfect cancellation is achieved if the following relation is satisfied:
\begin{align}\label{eq:strayfield}
    0 = \underset{\Delta \mathbf{I}_s}{\text{min}} \, 
    |\mathbf{B}_{\text{stray}} + B_s \Delta \mathbf{I}_s | \,.
\end{align}
In that case, the attainable minimum shape errors are:
\begin{align}
    \min_{\mathbf{I}_{e}} d(\mathbf{s}, \mathbf{s}^{\text{targ}}) = \min_{\mathbf{I}_s} d(\mathbf{s}, \mathbf{s}^{\text{targ}})\quad \forall \mathbf{I}_v, \mathbf{I}_o \,.
\end{align}
Thus, it would be sufficient to solve the following simplified minimization problem at every time step:
\begin{align}
    \min_{\mathbf{I}_s} d(S(F(\mathbf{I}_{e}, \bm{\theta})), \mathbf{s}^{\text{targ}})
    \equiv 
    IGS(\mathbf{s}^{\text{targ}}, \bm{\theta})  \,.
\end{align}
We will refer to this as the inverse Grad-Shafranov (IGS) problem. A number of different solvers address problems analogous or similar to the one stated here~\cite{hofmann1988fbt, faugeras2020overview, wai2026feedforward}.

\subsection{Practical considerations}\label{subsec:practical_controller}
The previous subsection showed that, under a set of simplifying assumptions, solving a time-independent IGS problem at every time step is in principle sufficient to solve the time-dependent problem stated in Eq.~\eqref{eq:mdp}. The main aim of this work is to develop a practical procedure that implements such solver in a real-time compatible manner, showing its suitability to achieve improved shaping capabilities in real-world tokamak experiments compared to standard pre-shot coil current optimization.
%
The problem of implementing a real-time compatible IGS solver is addressed by training a neural network surrogate of an existing IGS solver, which we denote $IGS_{nn}$. One challenge is that adding $\mathbf{I}_v$ to the input space of the neural network surrogate greatly expands the dimensionality of the input space that needs to be learned, which we found to be impractical given available resources. An estimate of the parameters $\bm{\theta}$ and of the vessel currents $\mathbf{I}_{v}$ can usually be obtained through an equilibrium reconstruction code. At TCV, this is done with the LIUQE code~\cite{moret2015tokamak}. This code also solves an IGS problem, but in this case the equilibrium and parameters result from a best fit to the available magnetic measurements rather than to targets imposed by the user. This is also the same code used in Sec.~\ref{sec:experiments} to evaluate the performance of the proposed method in experiments.

\begin{figure}
    \centering
    \includegraphics[width=\linewidth]{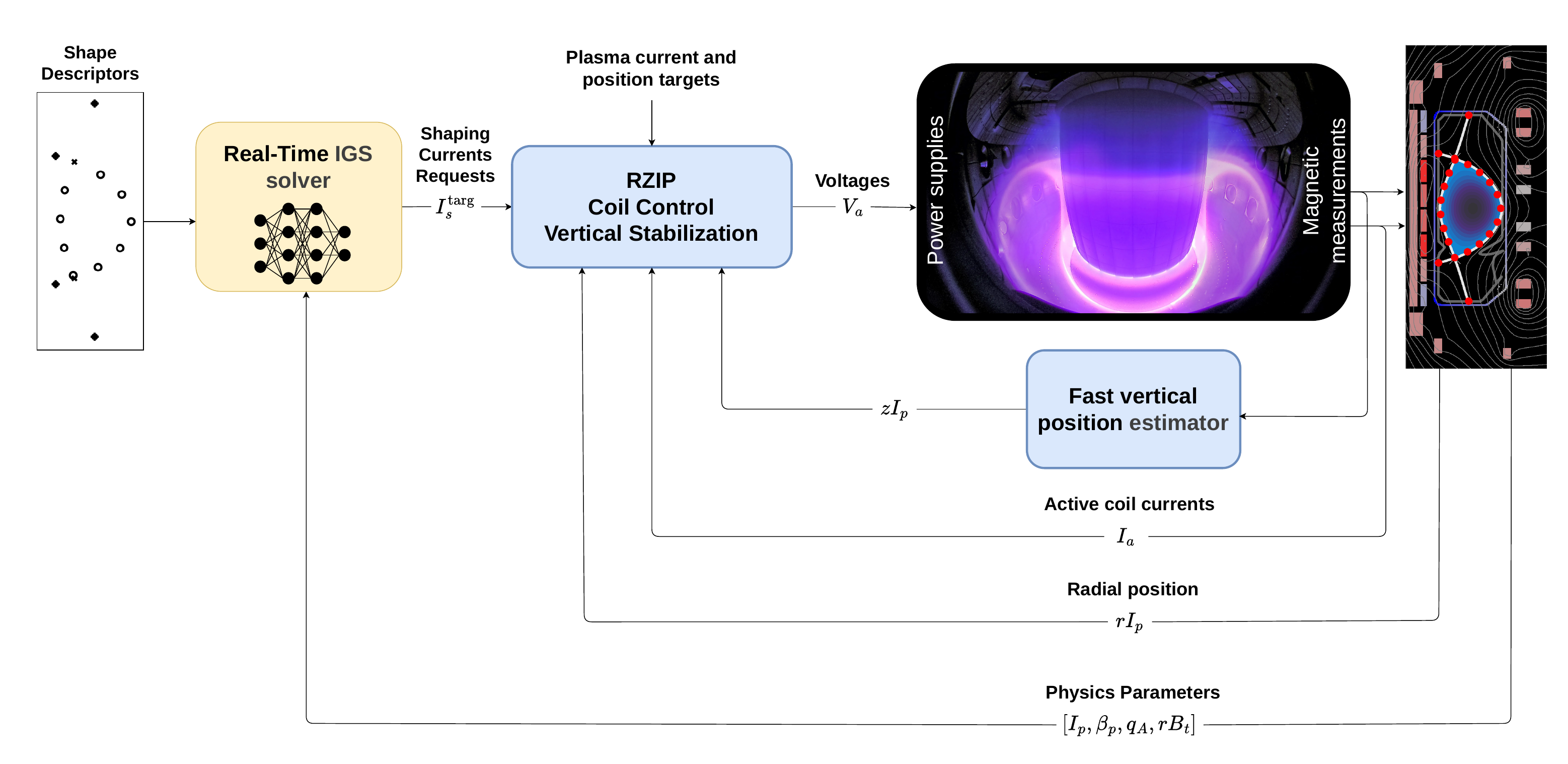}
    \caption{Simplified view of the shape control architecture including the proposed $IGS_{nn}$ solver in the case of the TCV tokamak. A set of shape descriptors is provided to the IGS solver, along with a reconstruction of a set of internal plasma parameters of interest. The proposed solver produces target values for the shaping coils currents $\mathbf{I}^{\text{targ}}_s$. These are passed to the lower-level control architecture in charge of regulating the active coil currents, the plasma position and current, and of vertically stabilizing the plasma; note that the current requests may be corrected with terms aimed at compensating the stray field, see Eq.~\eqref{eq:stray_cancel}. The resulting control voltages $\mathbf{V}_a$ are applied to the coils of the tokamaks. The magnetic measurements from the tokamak (magnetic fluxes and fields) together with the measured active currents $\mathbf{I}_a$ are fed to the RT equilibrium reconstruction algorithm (LIUQE in the case of TCV), which provides a full flux map and an estimate of the internal plasma parameters, which is fed back to the IGS solver. Assuming that the shaping currents are actuated perfectly, i.e. $\mathbf{I}_s \approx \mathbf{I}^{\text{targ}}_s$, the IGS solver is an approximate inversion of the plant including the equilibrium reconstruction, and the whole loop is approximately unitary.}
    \label{fig:full-loop}
\end{figure}

To make the proposed approach more general, $\mathbf{I}_o$ and $\mathbf{I}_v$ were not included among the inputs of $IGS_{nn}$, but a separate feedback cancellation term~\cite{underactuated} was employed instead to approximately compensate a generic poloidal stray field. This allows other effects, such as the presence of ferromagnetic materials, to be included in the proposed approach in a straightforward manner.

Note that, usually, perfect cancellation as in Eq.~\eqref{eq:strayfield} cannot be achieved in practice. However, this problem can be partially addressed by solving the following least squares problem: 
\begin{align}
    -B_s^{\dagger}\mathbf{B}_{\text{stray}} = \underset{\Delta \mathbf{I}_s}{\text{argmin}} \, 
        \left\lVert \mathbf{B}_{\text{stray}} + B_s \Delta \mathbf{I}_s \right\rVert_2 \,,
\end{align}
where $\dagger$ denotes the Moore-Penrose pseudo-inverse. From a practical point of view, since the full pseudo-inverse can lead to undesirably large coil current adjustments, a truncated pseudo-inverse retaining only the $n$ largest singular values can be employed, denoted $\dagger_n$. The choice of $n$ is a trade-off between effective stray field cancellation and acceptable coil current adjustments. The feedback cancellation term is thus:
\begin{align}\label{eq:stray_cancel}
    \Delta \mathbf{I}_{s, fbc}  \equiv -B_s^{\dagger_n}\mathbf{B}_{\text{stray}} = \underset{\Delta \mathbf{I}_s}{\text{argmin}} \, 
        \left\lVert \mathbf{B}_{\text{stray}} + B_s \Delta \mathbf{I}_s \right\rVert_2
\end{align}

In addition, the coupled tokamak-plasma system can have high frequency dynamics, e.g. vertical stability and passive conductor dynamics. The high frequency dynamics can be faster than the control response of the coils and even the cycle time of a neural network surrogate of the IGS solver. Thus, it is desirable to have high frequency feedback control on quantities with fast dynamics like the radial and vertical position of the plasma. The classical RZIP controller, which has seen widespread use in tokamak operation, can be used to address the coupled problem of shaping coil current, radial and vertical position, and plasma current control. The high frequency dynamics of passive conductors poses a challenge to the feedback cancellation approach. In principle, a low-pass filter on $\mathbf{I}_v$ can be employed to cancel out the low frequency content only, but we found such content to have a small impact for TCV in simulations, and thus neglected it in practice. 
The resulting control policy is:
\begin{subequations}\label{eq:final_policy}
\begin{empheq}[box=\fbox]{align}
    \mathbf{I}^{\text{targ}}_{\text{s}} &= \operatorname{IGS}_{\text{nn}}(\mathbf{s}^{\text{targ}}, \bm{\theta}) - B_s^{\dagger_n}B_{o}\mathbf{I}_{o} \label{eq:policy_coil_current}\\
    \mathbf{V}_a &= \operatorname{RZIP}(\mathbf{o}_{diag}, r^{\text{ref}}, z^{\text{ref}}, I_{p}^\text{ref}) + \operatorname{CURRENT\_CTRL}(\mathbf{I}^{\text{targ}}_s,\mathbf{I}_s)
\end{empheq}
\end{subequations}
where a dedicated controller, used to actuate the desired shaping currents $\mathbf{I}^{\text{targ}}_s$, is explicitly introduced along with the RZIP controller employed for plasma position and current. Note that typically this controller operates in a space that is orthogonal to the PF current directions adopted for radial and vertical position control.
Magnetic diagnostics used for RZIP control are denoted by $\mathbf{o}_{diag}$, while $r^{\text{ref}}$ and $z^{\text{ref}}$ are obtained from the shape descriptors, and $I_p^{\text{ref}}$ is the reference plasma current.

\subsection{Comparison with Reinforcement Learning control}\label{subsec:rl_connection}
Previous works on applying RL, namely policy optimization methods, to magnetic control have gained significant attention in the community~\cite{degrave2022magnetic,tracey2024towards}. 
%
To draw a connection between the approach proposed in this work and RL ones, we observe that the computational problem stated in Eq.~\eqref{eq:mdp} is equivalent to an MDP, which is one of the standard frameworks for formalizing RL problems. Cited works regarding the application of RL to tokamak magnetic control involved learning a control policy, $\pi$, which has diagnostic measurements as inputs. This requires the addition of an observation function, $O(\mathbf{I}_e, \bm{\theta})$, to generate synthetic diagnostic observations, $\mathbf{o}$ from the system's state and additional parameters. The resulting problem is a Partially Observable MDP (POMDP):
\begin{subequations}\label{eq:pomdp}
    \begin{align}
        \min_{\pi} &\sum_{t=1}^N d(\mathbf{s}_t, \mathbf{s}^{\text{targ}}_t)\\
        s.t. \quad&\mathbf{s}_t = S(F(\mathbf{I}_{e, t},\bm{\theta}_t))\\
        & \mathbf{I}_{e,t+1} \text{ per Eq.~\eqref{eq:circuit_dic}}\\
        & \mathbf{o}_t = O(\mathbf{I}_{e,t}, \bm{\theta}_t)\\
        & \mathbf{V}_{a, t} = \pi(\mathbf{o}_t)
    \end{align}
\end{subequations}
RL policy optimization applied to magnetic control can be viewed as an approximation algorithm to the POMDP stated in Eq.~\eqref{eq:pomdp}, where $\pi$ is a neural network. The control policy proposed in this work is an approximation of the MDP in Eq.~\eqref{eq:mdp}, under a certain set of assumptions. It can also be viewed as an approximation of the POMDP in Eq.~\eqref{eq:pomdp}, with physics-based equilibrium reconstruction methods used to estimate the state given observations. 

From this perspective, both the proposed control policy and RL policy optimization methods are approximation techniques to the POMDP stated in Eq.~\eqref{eq:pomdp}. RL policy optimization methods provide the largest degree of generality; extensions to the physical model can be handled naturally without any changes to the algorithm. The proposed control policy instead exploits structure in the POMDP, and as a consequence is less general. However, we argue it provides some interesting practical advantages. First, the classical RZIP controller provides a layer of separation between the neural network and the hardware, which in practice enables the implementation of an additional machine protection layer. Another advantage is that the neural network is a surrogate of a purely time-independent problem, whose training is significantly more efficient from a computational point of view than RL methods, which require time-dependent simulations. Finally, given that the neural network surrogates a physics-based procedure, sufficient advances in real-time IGS solvers might eventually obviate the need for a real-time neural network altogether. This feature in particular may be desirable in the context of burning plasma tokamaks, which are particularly sensitive to safety issues.

\clearpage

\section{Policy implementation}
This section describes the controller implemented in simulation and experiment in more detail. Sec.~\ref{subsec:dataset} defines the dataset generation process for training the neural network. Sec.~\ref{subsec:network_arch} details the neural network architecture implemented in this work. Sec.~\ref{subsec:training_details} provides further details on the model training process. As users rarely define shape targets for every time step, e.g. one millisecond time steps are typical, a simple shape target interpolation scheme was implemented and is described in Sec.~\ref{subsec:target_interpolation}.

Additional implementation details of lesser general interest, but of potential importance for re-implementation, are detailed in Appendix~\ref{app:additional_implementation}. {Namely, Sec.~\ref{subsec:voltage_ff} details a voltage feed-forward model for generating updates to voltage feed-forwards in real-time}, Sec.~\ref{subsec:10khz_control} details the integration of the $IGS_{nn}$ with coil current, radial and vertical position, and plasma current control, and Sec.~\ref{subsec:handover} details the handover strategy to transition from the breakdown and early ramp-up controller to the $IGS_{nn}$-based one.

\subsection{Dataset Generation}\label{subsec:dataset}
The IGS solver used for experiments in this work is FBT \cite{hofmann1988fbt,felici2025tokamak}. Datasets used for training, validation and testing were generated with large-scale runs of FBT through two workflows. The first involves using synthetic plasma shapes produced using the standard TCV pulse preparation workflow. For each shape in the pulse preparation workflow, sweeps on $\beta_p$, $I_p$, and $q_A$ were performed. For the real-time adaptation demonstrations detailed in Sec.~\ref{subsec:real_time_adaptation_demo}, a greater diversity of DN shapes was desired. Thus, the second workflow involved a parametric scan of possible DN shapes in addition to scans on plasma parameters. The design of the dataset used for training was adjusted over successive iterations along with the structure of the neural network. 
{In particular, the dataset for the initial experimental demonstration in Sec.~\ref{subsec:initial_demo} includes $\approx 3.6\times 10^5$ equilibria. The dataset used for simulations and experiments in the rest of the results section includes $\approx 2.2 \times 10^6$ equilibria spanning a diversity of plasma shapes. Table~\ref{tab:parameter_bound_dn_sweep} details both the physics parameters and DN shape parameters scanned for dataset generation.}

\begin{table}[!htbp]
\centering
\begin{tabular}{ccl}
\hline \textbf{Lower Bound} & \textbf{Upper Bound} & \textbf{Description} \\
\hline
 0.63 & 1.1 & Lower strike point r position [m] (only diverted plasmas) \\
 0.73 & 0.8 & Lower X-point r position [m] (only diverted plasmas) \\
 1.3 & 1.7 & Elongation [-]\\
0.5 & 2.5 & Inner Gap [cm]\\
 0.5 & 2.5 & Outer Gap [cm]\\
 $-300$ & $-150 $ & Plasma Current (kA) \\
 0.1 & 1.0 & Poloidal Beta [-] \\
0.8 & 1.3 & Safety Factor on Axis [-] \\
\hline
\end{tabular}
\caption{Parameter space swept for data generation.}
\label{tab:parameter_bound_dn_sweep}
\end{table}

\subsection{Network Architecture}\label{subsec:network_arch}
\begin{figure}[H]
    \centering
    \includegraphics[width=\linewidth]{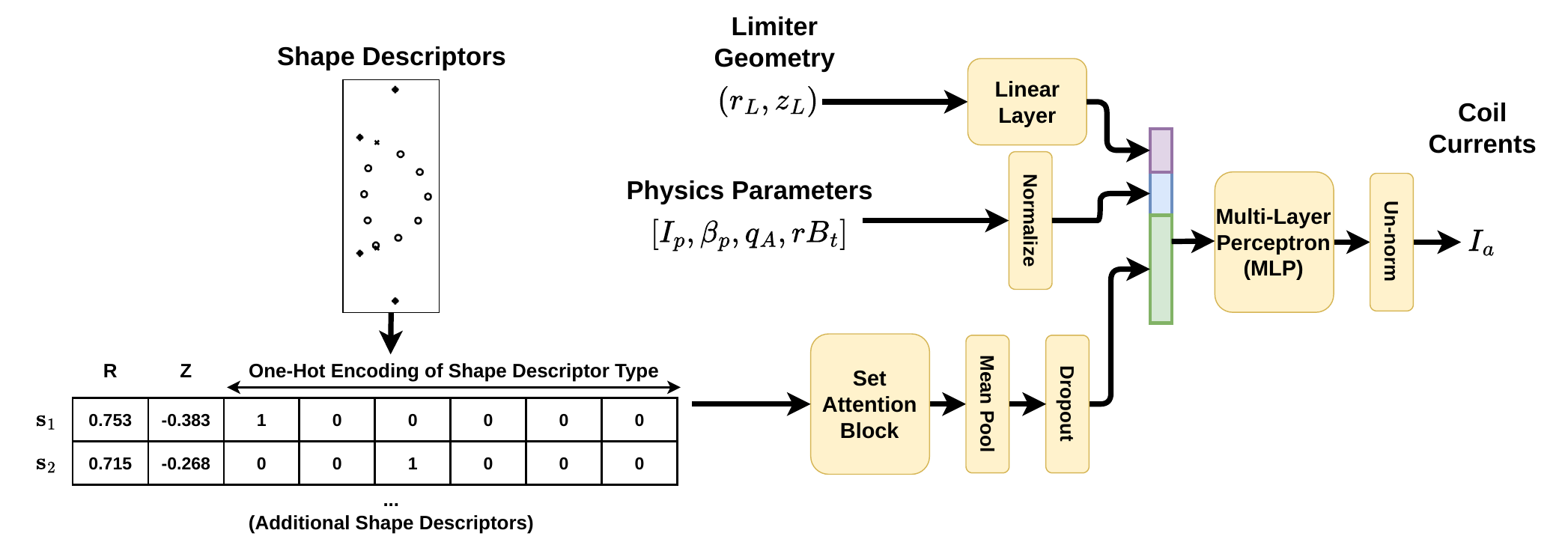}
    \caption{The architecture of the neural network surrogate of FBT deployed to the TCV plasma control system (PCS).}
    \label{fig:network_architecture}
\end{figure}

The neural network surrogate of FBT deployed in the TCV plasma control system (PCS) is illustrated in Fig.~\ref{fig:network_architecture}. 
%
{
The control points are divided into different categories, summarized in Table~\ref{tab:shape_descriptor_types} and represented with a one-hot encoding. Of these, category 5 has been introduced to allow controlling the $dr_{sep}$ quantity, defined as the flux difference between the primary and secondary separatrix, typically computed at the outer midplane. Category 6 instead flags points that are ignored (the maximum number of points that can be handled by the controller is fixed).
The latter is concatenated with the $(r, z)$ position of the shape descriptor to obtain an eight-dimensional input vector (see Fig.~\ref{fig:network_architecture}).}
A set-attention block \cite{lee2019set} then encodes these shape descriptors into a fixed-length vector. This design choice was made to accommodate a variable number of shape descriptors and control points, which vary with the plasma configuration (e.g.\ limited vs.\ single null vs.\ DN). Moreover, it captures the input order invariance of the shape descriptors. The output of this set-attention block is concatenated with normalized scalar physics parameters ($I_p$, $\beta_p$, $q_A$, $rB_t$) and a scalar encoding of the limiter geometry. This encoding was obtained from a single linear layer applied to the concatenated $(r, z)$ coordinates of the limiter points, which varied over the experiments performed in this work due to the presence/absence of movable baffles in the vacuum chamber~\cite{reimerdes2021initial}. The combined vector is then passed through a multi-layer perceptron (MLP) with ReLU activations and un-normalized to yield the predicted active coil currents.

Hyperparameter sweeps were performed with the standard Bayesian optimization algorithm implemented by Weights and Biases~\cite{wandb}. The best model, according to the validation loss metric, was selected for each deployment. The sweep configuration is detailed in Table~\ref{tab:hyperparams}.

\begin{table}[h]
\centering
\caption{\label{tab:shape_descriptor_types}Shape descriptor types.}
\begin{tabular}{cl}
\hline\hline
Index & Description \\
\hline
1 & Boundary control point \\
2 & Strike point \\
3 & Primary X-point \\
4 & Secondary X-point \\
5 & Secondary separatrix point, used to compute $dr_{sep}$ \\
6 & Ignored point \\
\hline\hline
\end{tabular}
\end{table}

\begin{table}[h]
\centering
\caption{\label{tab:hyperparams}Hyperparameter sweep configuration.}
\begin{tabular}{lll}
\hline\hline
Parameter & Type & Range \\
\hline
MLP width & Discrete & \{128, 256\} \\
MLP depth & Discrete & \{2, 3, 4\} \\
Shape descriptor encoding dimension & Integer & [10, 18] \\
Attention heads & Integer & [4, 8] \\
Scaling of input into the set attention block & Continuous & [0.5, 2.0] \\
Dropout rate & Continuous & [0.0, 0.5] \\
\hline\hline
\end{tabular}
\end{table}

\subsection{Training Details}\label{subsec:training_details}
The training dataset consists of the FBT runs described in Sec.~\ref{subsec:dataset}, split 70/15/15 into train, validation, and test sets. The batch size was 16,384 samples. Training proceeded for a maximum of 5,000 epochs, with validation evaluated every 10 epochs; the best-performing model on the validation set was selected for deployment.

\subsubsection{Normalization}
The scalar physics inputs ($I_p$, $\beta_p$, $q_A$, $rB_t$) and the output active coil currents are each normalized by subtracting the training-set sample mean and dividing by the difference between the 75th and 25th quartiles. The $(r, z)$ coordinates of the shape descriptor control points are left unnormalized, as they are naturally close to unity.

\subsubsection{Loss function}
The loss is the mean Huber loss on the normalized prediction error across all active coils
\begin{align}
    \mathcal{L} = \frac{1}{N_c}\sum_{i=1}^{N_c} \ell_\delta\!\left(\frac{I_{a,i}^{\mathrm{pred}} - I_{a,i}^{\mathrm{targ}}}{I^{\mathrm{norm}}}\right),
\end{align}
where $N_c$ is the number of active coils, $I^{\mathrm{norm}} = 7500\,\mathrm{A}$ is a fixed normalization factor, and $\ell_\delta$ is the Huber loss with threshold $\delta = 0.1$. The Huber loss is used rather than the mean-squared error to reduce sensitivity to occasional outlier equilibria in the training data.

\subsubsection{Optimizer}
The AdamW optimizer~\cite{loshchilov2017decoupled} is used with weight decay $\lambda = 5\times10^{-3}$. The learning rate follows a warm-up--cosine-decay schedule: linearly warming up from $5\times10^{-4}$ to a peak value of $1\times10^{-2}$ over 100 gradient steps, then decaying via a cosine schedule to a final value of $5\times10^{-3}$ over 20,000 gradient steps. Dropout with rate 0.33 is applied to the set attention output during training.

\subsubsection{Model Performance}
\begin{figure}[!htbp]
    \centering
    \includegraphics[width=0.7\linewidth]{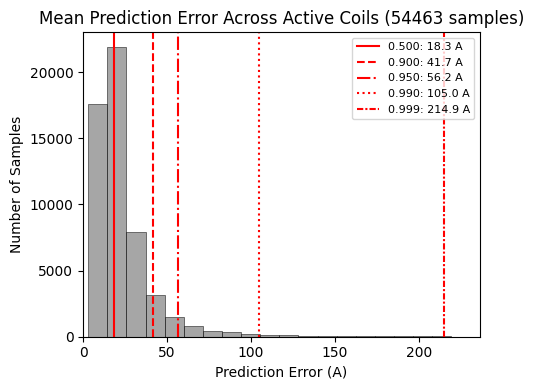}
    \caption{Histogram of prediction error of the neural network used in TCV\#86713 on the test set. 
    Vertical lines corresponding to select quantiles are also shown.}
    \label{fig:network_for_87970}
\end{figure}

Coil current prediction errors of $\sim56$ and $\sim105$ Amperes per coil were found to be achievable for 95\% and 99\% of cases in the test set respectively, with {median} prediction errors of $\sim 18$ Amperes per coil, as shown in Fig.~\ref{fig:network_for_87970}. This is notable as few tens of Amperes of coil current measurement error are considered acceptable for TCV experiments, giving confidence that the network's error is within a reasonable range. 

\subsection{Interpolating Targets}\label{subsec:target_interpolation}
In practice, shape targets are usually not programmed with a millisecond time-resolution; at TCV, usually a set of 10-50 target shapes are specified across the pulse. The matter of optimally specifying a fine-grained shape target trajectory is a separate problem outside the scope of this paper~\cite{wai2026feedforward}. In this work, a simple heuristic strategy is adopted. Consider two shape targets $\mathbf{s}^{\text{targ}}_i$ and $\mathbf{s}^{\text{targ}}_{i+1}$ corresponding to times $t_i$ and $t_{i+1}$. If the two shape targets have the same structure, e.g. they are both DN equilibria described by analogous sets of control points, then the two vectors are linearly interpolated to obtain the intermediate shape targets:
\begin{align}
    \mathbf{s}^{\text{targ}}(t) = \mathbf{s}^{\text{targ}}_i + \frac{t - t_i}{t_{i+1} - t_i}(\mathbf{s}^{\text{targ}}_{i+1} - \mathbf{s}^{\text{targ}}_i), \quad t \in [t_i, t_{i+1}]
\end{align}
Note that this is different from conventional approaches, which typically resort to interpolating coil currents between target shapes. This is what is done, for example, in the standard TCV preparation procedure.
The proposed $IGS_{nn}$ allows direct interpolation of the shape targets instead, providing a finer control over the desired plasma shape during transient phases. However, in the present implementation this is limited to cases where the description of the plasma shape is consistent between consecutive snapshots in terms of both number and type (strike, null, boundary, etc.) of contour points. If two adjacent shape targets have different structures, e.g. one is limited and the other is diverted, then the interpolation is still performed on the coil current outputs produced from $IGS_{nn}$.

\section{Experimental and Simulation Results}\label{sec:experiments}
The developed control strategy has been tested in closed-loop simulations with the FGE code~\cite{heiss2025fge, carpanese2021development} and deployed in the PCS of TCV. However, due to time constraints and unforeseen hardware issues, not all developed capabilities and scenarios were experimentally tested during the experimental campaign. Thus we present results obtained from both simulations and TCV experiments with further improvements for experimental deployment being left for future work.
To clearly distinguish between simulations and experimental results, we plot the reconstructed experimental equilibria with a dark background, while simulated plasma snapshots feature a clear one.

A summary of the experiments discussed in this section is given in Table~\ref{tab:experiments}.

\begin{table}[htbp]
\centering
\caption{Summary of TCV shots referenced in this work, the scenario each was used to demonstrate, and the section(s) in which it appears.}
\label{tab:experiments}
\begin{tabular}{lll}
\hline
Shot(s) & Scenario / Controller & Section(s) \\
\hline
\#86708, \#86713 & LSN, NBH-induced $\beta_p$ disturbance ($IGS_{nn}$)         & Sec.~IV A \\
\#87970          & Symmetric DN, strike point sweep ($IGS_{nn}$); & Sec.~IV B, IV C \\
                 & \quad also early-termination demonstration                  & \\
\#86710          & Symmetric DN, standard RZIP control (baseline)              & Sec.~IV B \\
\#86310          & Symmetric DN, 10~kHz DN feedback controller (baseline)~\cite{lafferty2026fast} & Sec.~IV B \\
\#89936          & Asymmetric DN, $IGS_{nn}$ control                           & Sec.~IV B \\
\#79114          & Asymmetric DN, isoflux control (baseline)~\cite{mele2025design} & Sec.~IV B \\
\#79102          & Asymmetric DN, standard RZIP control (baseline)             & Sec.~IV B \\
\hline
\end{tabular}
\end{table}

\subsection{Single Null scenario}\label{subsec:initial_demo}
\begin{figure}[!htbp]
    \centering
    \includegraphics[width=\linewidth]{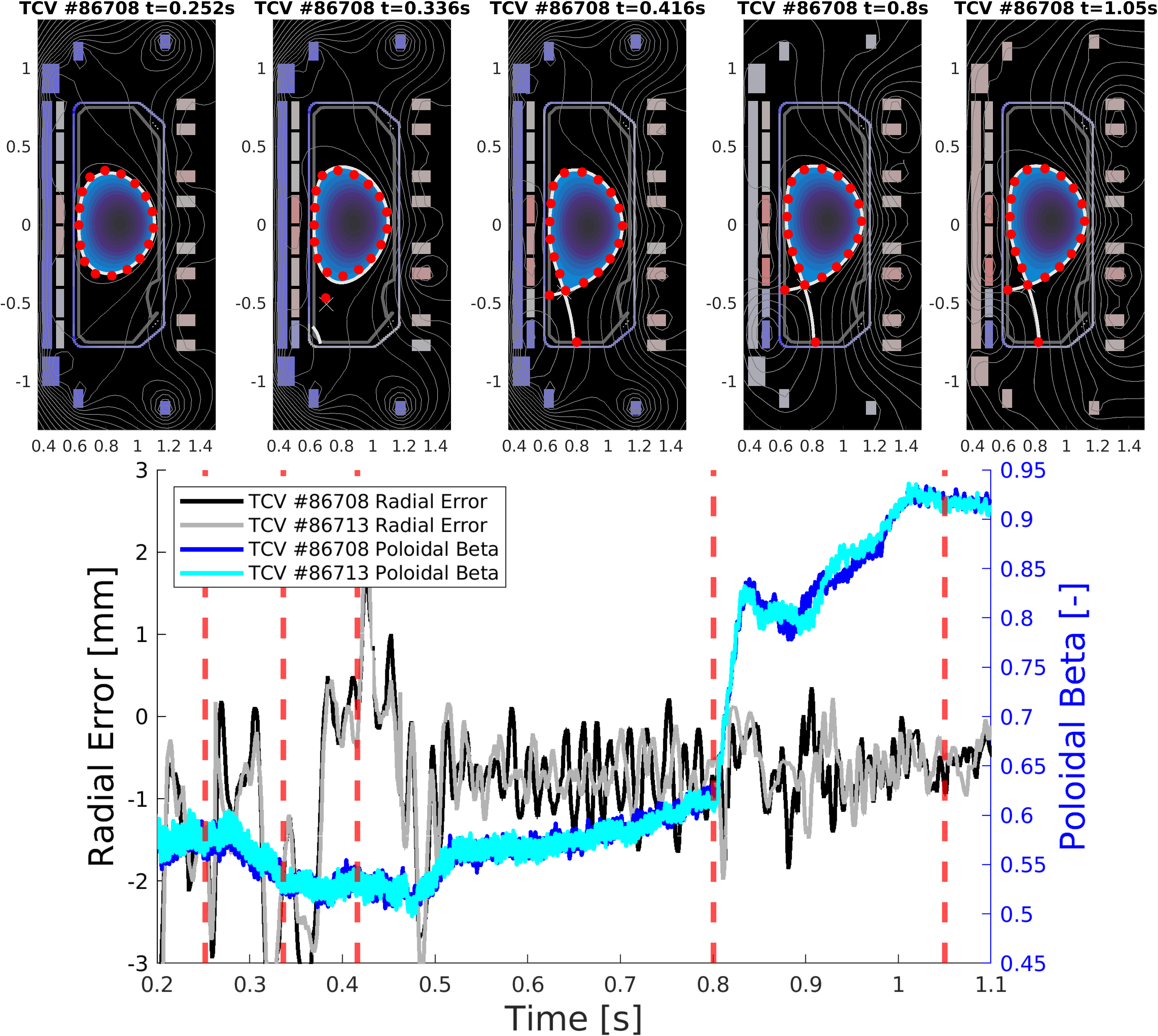}
    \caption{(Top) equilibrium reconstruction at select time slices of TCV \#86708, with the next shape target visualized with red dots. (Bottom) the poloidal beta, in shades of blue, and radial control error, in shades of gray, for the two pulses.}
    \label{fig:lsn_result}
\end{figure}

Initial experimental demonstration involved a LSN scenario, with Neutral Beam Heating (NBH) added to introduce a pressure disturbance. This scenario was successfully executed twice in TCV shots \#86708 and \#86713, with the results summarized in Fig.~\ref{fig:lsn_result}. Both pulses feature millimeter precision in radial control during the diverted phase, with minimal disturbance of the overall equilibrium despite the introduction of NBH.

\begin{figure}[!htbp]
\centering
\includegraphics[width=0.7\linewidth]{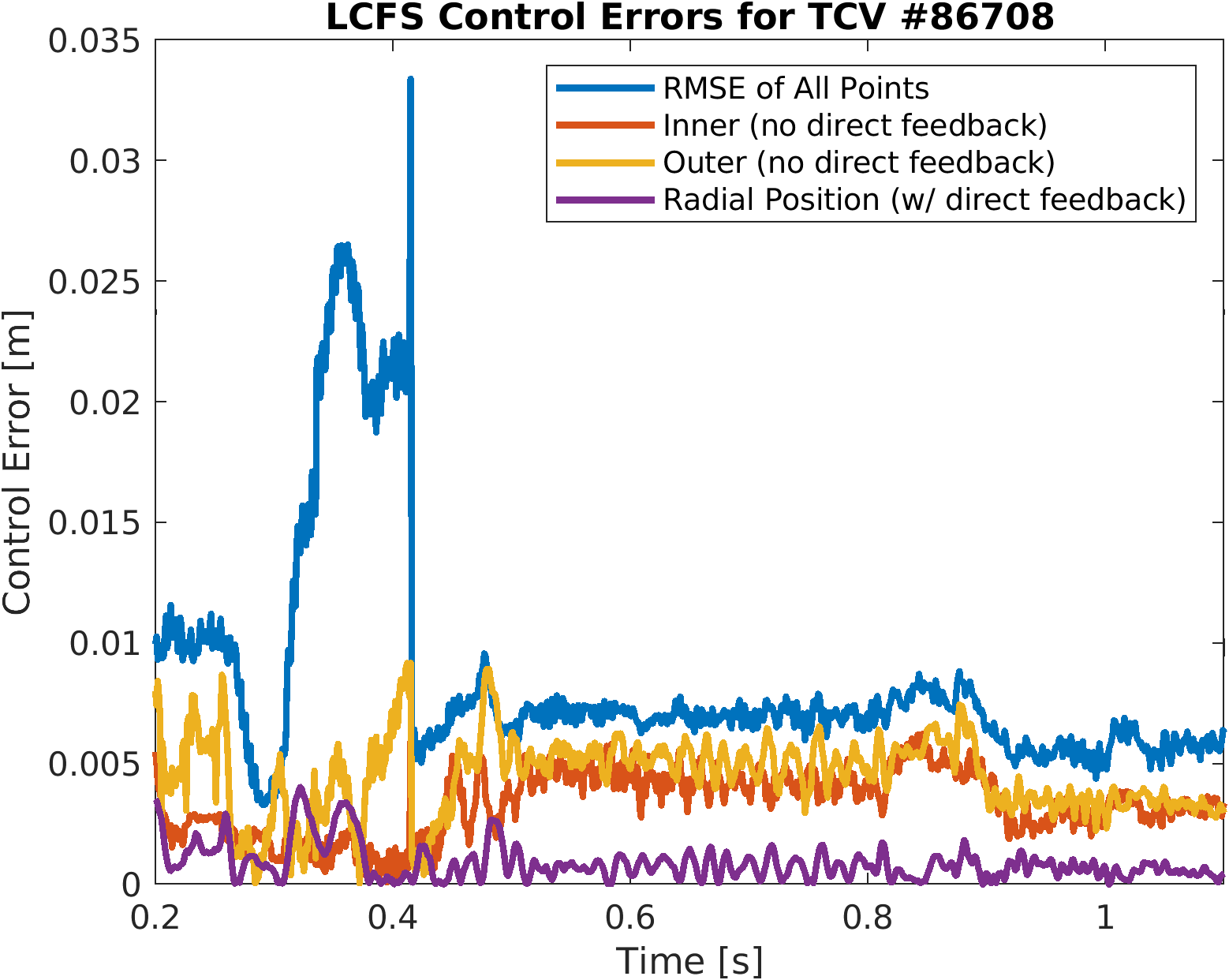}
\caption{Control errors for the radial position, inner and outer most control points, and the RMSE of all control points for TCV \#86708.}
\label{fig:lcfs_errors_86708}
\end{figure}

Additional control error metrics are plotted in Fig.~\ref{fig:lcfs_errors_86708}. 
{Note that the seemingly high Root Mean Square Error (RMSE) of $\sim 2.5$ cm during the limiter to diverted transition is related to the interpolation strategy described in Sec.~\ref{subsec:target_interpolation}. In particular, since the type of control points is changing from one reference shape to the next, the interpolation is performed on the current outputs of $IGS_{nn}$, and the shape error is computed with respect to the \textit{next} shape target. This explains the sudden drop observed just after 0.4s, once the divertor is fully formed.}
In the diverted phase, the inner and outermost control points see errors of $\sim 5$mm, with an overall RMSE of $\sim 7$mm. The error appears to be systematic, indicating that the combination of errors in the control strategy (e.g. NN prediction error and coil current control precision) is non-negligible. 
Notably, the radial position, which has an additional feedback loop, saw a negligible $\sim 1$mm error, highlighting the potential value of integrating the developed control scheme with additional closed-loop shape feedback. Nevertheless, the observed control errors are below typical errors in equilibrium reconstruction, cited as $1$cm in~\cite{hommen2014real}. 
\clearpage

\subsection{Double Null with Strike Point Sweep}
The developed controller was applied to DN plasmas in experiment and compared against similar discharges employing different control strategies. Prior work on TCV has applied $1$kHz isoflux control to an up-down asymmetric DN~\cite{mele2025design} as well as a fast $10$kHz DN balancing controller to a symmetric DN~\cite{lafferty2026fast}. The DN imbalance is quantified in terms of flux difference between the two X-points, $\Delta \psi_X$; a zero flux difference nominally corresponds to perfect balancing.

\begin{figure}[!htbp]
    \centering
    \includegraphics[width=\linewidth]{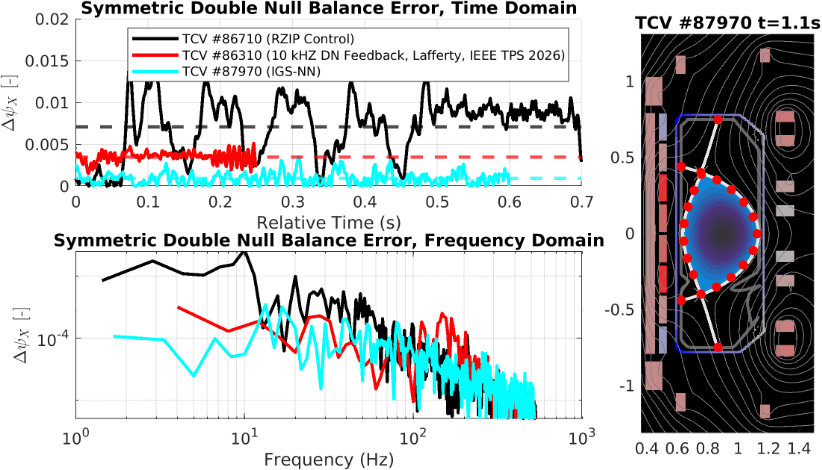}
    \caption{DN imbalance in the time (top left) and frequency (bottom left) domains for the up-down symmetric DN scenario. Time traces shown are relative to $0.5$s for \#86710, $0.55$s for \#86310, $0.6$s for \#87970. (Right) A time slice of the reconstructed equilibrium controlled with the $IGS_{nn}$ along with control targets (red dots).}
     \label{fig:dn_compare_symmetric}
\end{figure}

A demonstration of accurate DN balancing in the symmetric case was achieved with the $IGS_{nn}$ in TCV \#87970, which also featured strike point sweeping. 
Fig.~\ref{fig:dn_compare_symmetric} compares the precision achieved by the $IGS_{nn}$ approach (\#87970) with a pulse using standard RZIP control (\#86710) and a recent pulse utilizing $10$kHz linear control of the DN imbalance (\#86310)~\cite{lafferty2026fast}. In terms of mean time-domain error, the $IGS_{nn}$ controller shows clear improvements over both RZIP and the $10$kHz DN controller for this symmetric case. However, it should be noted that the fast $10$kHz scheme relied on a simplified flux observer, which might introduce a small systematic estimate error with respect to the nominal flux difference provided by full offline equilibrium reconstruction (used for this comparison). Moreover, most of the pulse in the $10$kHz linear controller involved vertical position shifts; thus, to provide a fair comparison, only the portion of the pulse without such shifts was considered.

\begin{figure}[!htbp]
    \centering
    \includegraphics[width=\linewidth]{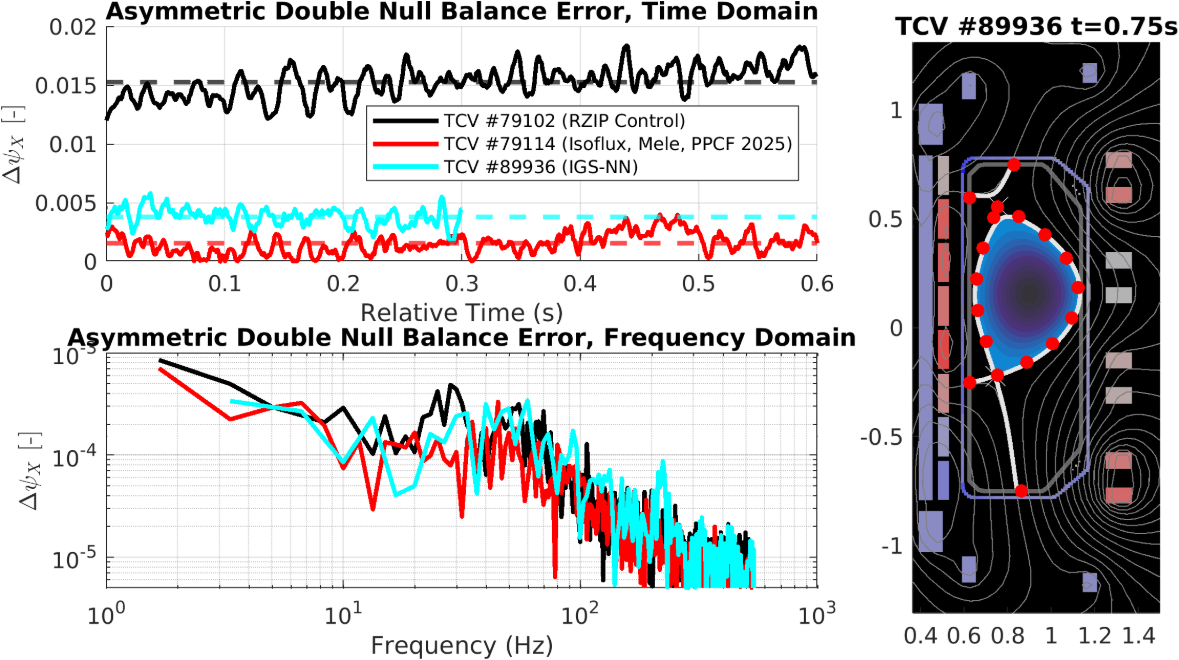}
    \caption{DN imbalance in the time (top left) and frequency (bottom left) domains for the up-down asymmetric DN scenario. Time traces shown are relative to $1.2$s for \#79102, $1.2$s for \#79114, and $0.7$s for \#89936. (Right) A time slice of the reconstructed equilibrium controlled with the $IGS_{nn}$ along with control targets (red dots).}
    \label{fig:dn_compare_asymmetric}
\end{figure}

A comparison was also performed for the asymmetric shape controlled with isoflux control in TCV \#79114. In pulse \#89936 (see Fig.~\ref{fig:dn_compare_asymmetric}) the $IGS_{nn}$ controller improves DN balance significantly relative to the RZIP baseline, but falls short of the performance of the isoflux controller. This performance degradation may be attributed to any combination of approximations and systematic biases in the control architecture (e.g. NN approximation error, RZIP control error, IGS modelling inaccuracies), and highlights again the potential desirability of incorporating additional feedback with the developed control scheme.

\clearpage
\subsection{Real-time Adaptation Capabilities}\label{subsec:real_time_adaptation_demo}
\begin{figure}[!htbp]
    \centering
    \includegraphics[width=0.9\textwidth]{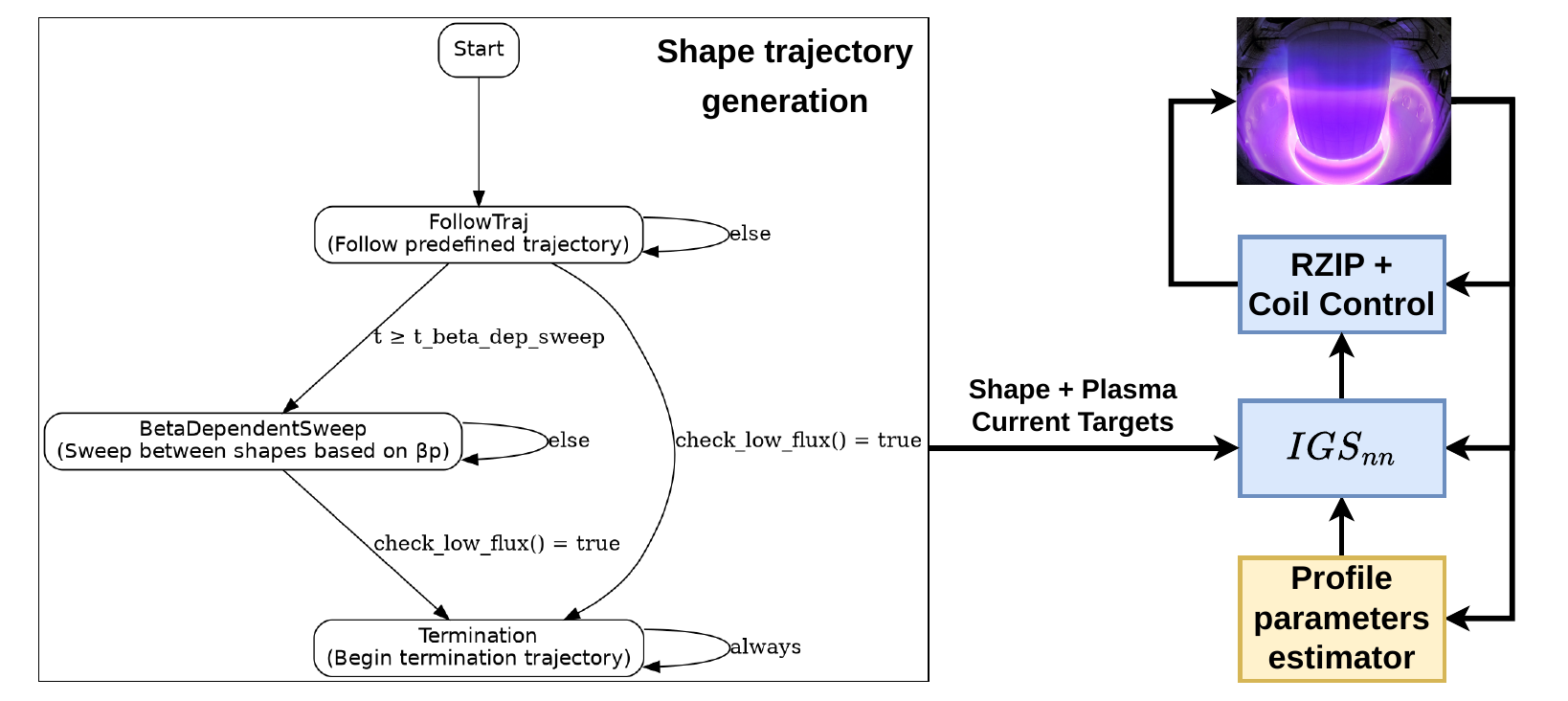}
    \caption{Depiction of the state machine used for the real-time adaptation demo, and how it integrates with the rest of the magnetic control system. By default, the state machine starts in the \texttt{FollowTraj} state, where a pre-defined shape trajectory is passed to the $IGS_{nn}$ solver. If the current time is larger than \texttt{t\_beta\_dep\_sweep}, the $\beta$-dependent strike points sweep is activated, meaning that the amplitude of the sweep will be adjusted proportionally to the estimated $\beta_p$. A further condition based on the remaining OH flux triggers another transition to the \texttt{Termination} state, which generates references for a controlled ramp-down. The generated references are passed to the $IGS_{nn}$ solver, which also takes direct feedback from the machine (e.g. coil current measurements) and estimates of relevant plasma internal profile parameters (e.g. $\beta_p, q_A$), generating references for the lower level coil currents and RZIP controllers.}
    \label{fig:state_machine}
\end{figure}

A further DN scenario was designed to demonstrate the real-time adaptation capabilities enabled by the proposed method. 
A simple state machine, depicted in Fig.~\ref{fig:state_machine}, was implemented to adapt the shape and plasma current targets sent to the $IGS_{nn}$ controller with the goal of:
\begin{enumerate}
    \item adjusting the magnitude of the strike point sweep to increase with measured $\beta_p$;
    \item initiating an early ramp-down once remaining Ohmic flux falls below a given threshold.
\end{enumerate}
As a proof-of-principle, the strike point sweep amplitude was adjusted based on $\beta_p$, given that a reliable estimator of this quantity is available in real-time at TCV. Low Ohmic flux was similarly chosen as a simple real-time trigger to demonstrate the ability to initiate a well-controlled early ramp-down.
A more comprehensive demonstration of machine protection capabilities might involve adapting the strike points -- and/or the overall shape -- in response to real-time estimates of quantities like heat flux, tile temperatures, or detachment state. Strategies along these lines have recently been proposed for ITER to manage heat loads~\cite{frattolillo2025magnetic,pesamosca2024development}. Similarly, alternative triggers might be chosen for early ramp-down initiation. The integration of the proposed method with more sophisticated machine monitoring and protection strategies is left for future research.

\begin{figure}[!htbp]
    \centering
    \includegraphics[width=0.8\textwidth]{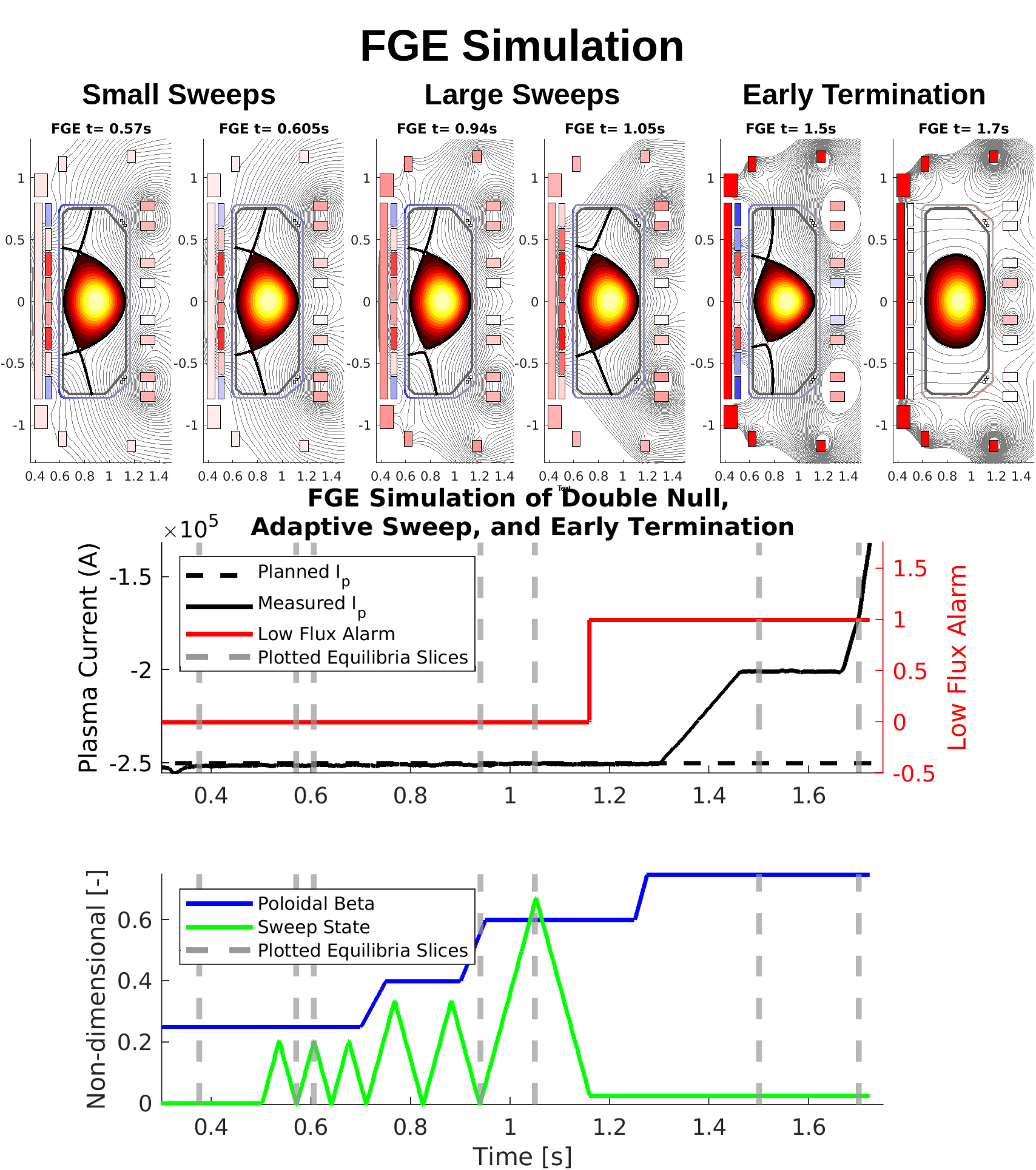}
    \caption{FGE simulation results demonstrating real-time adaptation capabilities of the developed method. (Top) equilibria visualized for selected time slices, illustrating varying strike point sweep amplitudes and early plasma termination. (Center) a plot of the pre-planned and measured plasma current, showing the early termination that is initiated once the low flux alarm is triggered. (Bottom) a plot of the poloidal beta along with the amplitude of the strike point sweep, which ranges between 0 and 1. Center and bottom plots also show the times corresponding to the visualized equilibria.}
    \label{fig:dn_adapt_scenario}
\end{figure}

This demonstration scenario was successfully realized in simulation, with partial success in experiment. Fig.~\ref{fig:dn_adapt_scenario} summarizes simulation results demonstrating real-time adaptation. In this simulation, $\beta_p$ was prescribed to increase, resulting in an increase in the strike point sweeping amplitude. Once the low flux alarm is triggered, a controlled early termination is initiated, reducing the plasma current and the plasma elongation with respect to the pre-planned trajectory before terminating the plasma.

\begin{figure}[!htbp]
    \centering
    \includegraphics[width=0.8\textwidth]{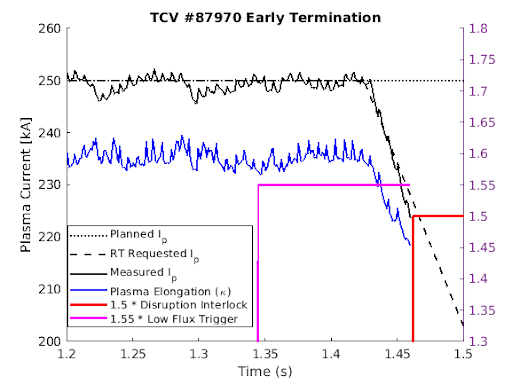}
    \caption{Experimental results from TCV \#87970 demonstrating early termination capabilities; boolean trigger signals are scaled for visualization on the shared axis. In response to the low flux alarm (pink), the state machine transitioned to the early termination trajectory, requesting a deviation of the plasma current trajectory (requested in dashed black, measured in solid black) away from the pre-planned trajectory (dotted black). The elongation is also decreased in a controlled fashion (blue). The machine protection system treated this excursion in the plasma current as a disruption, intervening and actually disrupting the plasma (red).}
    \label{fig:tcv_87970_early_term}
\end{figure}

Early termination was partially demonstrated in TCV pulse~\#87970, as shown in Fig.~\ref{fig:tcv_87970_early_term}. The low flux alarm successfully triggered a well-controlled ramp-down of the plasma current and elongation. However, the rapid change in plasma current triggered the machine protection systems disruption detector, which intervened by disabling hardware. This led to an actual disruption of the plasma. Similarly, adaptive strike point sweeping was also demonstrated only partially in experiments. The increase in $\beta_p$ observed in the discharge was lower than expected, potentially due to the presence of impurities in the vacuum chamber. As a consequence, the strike point sweep period remained approximately constant.
\clearpage

\subsection{One Network to Rule Them All}\label{subsec:diversity_shapes}
\begin{figure}[!htbp]
    \centering
    \includegraphics[width=0.9\textwidth]{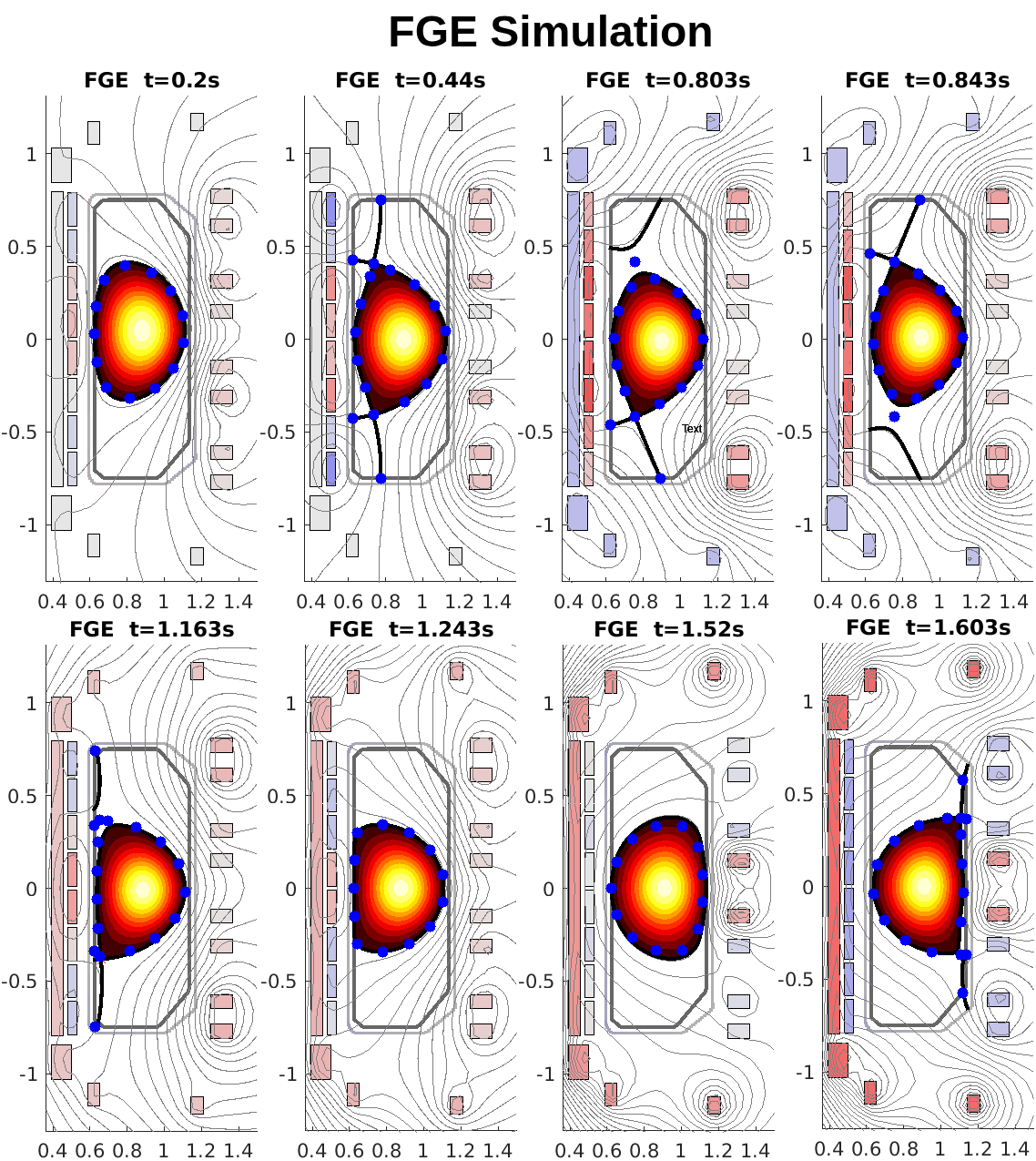}
    \caption{FGE simulation of a pulse which moves the plasma through a sequence of limited, single null, and DN shapes with both negative and positive triangularity. All shapes were controlled to targets (blue dots) with a single neural network, which was also used for the real-time adaptation demonstration in Sec.~\ref{subsec:real_time_adaptation_demo}.}
    \label{fig:plasma_storm_sim}
\end{figure}

An important feature of the control architecture is its ability to provide control of a diverse set of shapes. While it is in principle possible to achieve this feature with RL, existing approaches have not yet demonstrated this capability \cite{degrave2022magnetic,tracey2024towards,subbotin2025reconstruction}. It is much more computationally tractable to achieve this feature with the developed architecture, as the neural network training only needs to be done for the purely time-independent IGS problem. To demonstrate this feature, a single neural network was trained to control the shapes in both the real-time adaptation demonstration in Sec.~\ref{subsec:real_time_adaptation_demo} and also a pulse which moves the plasma through a sequence of diverse shapes shown in Fig.~\ref{fig:plasma_storm_sim}. This neural network was trained on a single A100 GPU in approximately 6 hours, which is considerably more compute-efficient than end-to-end RL based approaches, which reportedly involve tens to hundreds of parallel simulators running for $\sim$~10 hours to days~\cite{tracey2024towards,degrave2022magnetic}.

\clearpage
\section{Discussion}

\subsection{Limitations}
In this section, we briefly discuss some limitations of the proposed approach.

First of all, the integration between the $IGS_{nn}$ and RZIP and coil current controllers can be a source of practical challenges. In particular, the bandwidth of the coil current variations requested by the $IGS_{nn}$ should be explicitly limited to avoid undesired interactions between the two systems. For example, the radial component of the RZIP control may try to modify the coil currents in a direction that conflicts with $IGS_{nn}$ commands. 

{Similarly, the disruption observed during the early termination demonstration in Sec.~\ref{subsec:real_time_adaptation_demo} (Fig.~\ref{fig:tcv_87970_early_term}) was not caused by the developed controller itself. Rather, the disruption resulted from an interaction with another machine protection subsystem — the disruption interlock — which interpreted the rate of the commanded current change as indicative of an uncontrolled event, and intervened by disabling hardware. This highlights another practical integration challenge when deploying new control strategies alongside legacy protection/control systems, rather than a limitation of the proposed method \emph{per se}. Addressing this will likely require either adapting protection system thresholds to accommodate faster, intentional current ramp-downs, or coordinating planned trajectory changes with the protection logic in advance.}

Moreover, feedback cancellation of the vessel currents was successfully tested in simulation, but eventually it was not adopted in experiment. This was due to concerns with the real-time estimates of the vessel currents, which can be affected by noise and artifacts that might lead to undesirable control adjustments. This, in practice, limits the validity of the assumptions on which the optimality of the method is based.

Finally, while the developed approach improves on pure feedforward coil current optimization (in particular thanks to the availability of a real-time estimate of parameters related to the internal plasma current distribution) and is capable of adapting to asynchronous events, 
it does not employ explicit feedback on error terms. This means, in practice, that its robustness against disturbances and offsets is limited compared to closed-loop solutions. To ameliorate this issue, integration of closed-loop shape control corrections with the proposed approach could be considered. 

\subsection{Implications for Power Plants}\label{subsec:power_plant_implications}
Beyond the commissioning phase, the high neutron fluence in future power plants is expected to restrict the usable diagnostics to those that can operate behind sufficient neutron shielding. This is particularly problematic for conventional approaches to magnetic control, which depend on accurate plasma shape estimation.

This work provides a possible path to ameliorating this issue by showing that online optimization of coil currents given observed bulk plasma statistics can help correct for shape errors. If these plasma statistics can be observed with power plant compatible diagnostics, the developed approach may provide an accurate way of achieving the desired plasma configuration in a power plant environment. 


\subsection{Accounting for Additional Constraints}
While the networks deployed in the experiments described in this article did not explicitly account for constraints and loss terms in the IGS problem, such as coil current limits and structural loading, these could be included within the presented framework. One potential way of doing so would be adding input modality to the network and expanding the training dataset to sweep the parameter range of interest. This extension could prove especially relevant for usage in routine operations, as constraints and loss terms may be adjusted between shots.



\section{Conclusion}
This paper presents a new approach to tokamak magnetic control. The control strategy is motivated by a theoretical analysis of the magnetic control problem from the perspective of MDPs and POMDPs. The theoretical analysis also draws connections to deep RL, showing both deep RL and the developed control policy can be viewed as approximations to a POMDP defining the magnetic control problem.

Experiments at TCV show precise plasma shaping, with improved flexibility upon customary pre-shot planning tools and heavily model-based control solutions. Further simulation demonstrations of real-time adaptation and control of a diverse set of shapes with a single neural network further illustrate the benefit of the developed approach.


Future work should aim to deploy the developed control strategy in routine operations, which would yield further practical experience of the benefits and drawbacks of the approach. The developed approach also points to the possibility of achieving accurate magnetic control without in-vessel magnetic diagnostics, which future work should also pursue.

\section{Author Contributions}
Allen Wang conceptualized, developed, and implemented the controller. Adriano Mele led the overall deployment and integration of the controller into TCV and developed a significant part of the controller testing infrastructure. Cosmas Hei\ss{} developed the underlying 10kHz fast magnetic controller and contributed to experimental integration, testing, and debug. Cristian Galperti led the overall software integration and debug in the real-time control system. Zander Keith helped develop the machine learning pipeline used in this work. Alessandro Pau helped setup scenarios for initial testing. Antoine Merle helped with integration of real-time equilibrium reconstruction. Olivier Sauter advocated for the project and contributed to discussions that led to the key insights of this work. Daniel Gonzalez helped with software deployment and debugging. Francesco Carpanese helped with FGE simulations. Federico Felici pointed out the need to account for the Ohmic coil fields, helped resolve issues with equilibrium codes, and contributed to discussions that led to the key insights of this work. Mark Dan Boyer helped motivate the control strategy and contributed to discussions that led to the key insights of this work. Cristina Rea managed the collaboration and contributed to discussions that led to the key insights of this work.

\section{Acknowledgements}
This work has been carried out within the framework of the EUROfusion Consortium, funded by the European Union via the Euratom Research and Training Programme (Grant Agreement No 101052200 — EUROfusion). Views and opinions expressed are however those of the author(s) only and do not necessarily reflect those of the European Union or the European Commission. Neither the European Union nor the European Commission can be held responsible for them.
This work was funded in part by Commonwealth Fusion Systems.

\bibliography{apssamp}

\clearpage
\onecolumngrid
\appendix

\section{Additional Implementation Details}\label{app:additional_implementation}

\subsection{Voltage Feed-forward Model}\label{subsec:voltage_ff}
As coil currents are adjusted in real-time, it is important to re-compute voltage feedforwards to achieve the desired current trajectory. In addition, plasma current must also be regulated in real-time, and this task is coupled with the one stated above. To tackle this issue, the following coupled circuit equations are employed for computing the voltage feed-forwards:
\begin{subequations}
    \begin{align}
        M_{aa}\dot{\mathbf{I}}_a + R_a \mathbf{I}_a + M_{ao}\dot{\mathbf{I}}_o + M_{av}\dot{\mathbf{I}}_v + M_{ap} \dot{I}_p &= \mathbf{V}_a \label{eq:aa}\\
        M_{oo}\dot{\mathbf{I}}_o + R_o \mathbf{I}_o + M_{oa}\dot{\mathbf{I}}_a + M_{ov}\dot{\mathbf{I}}_v + M_{op} \dot{I}_p &= \mathbf{V}_o \label{eq:oo}\\
        M_{vv}\dot{\mathbf{I}}_v + R_v \mathbf{I}_v + M_{va}\dot{\mathbf{I}}_a + M_{vo}\dot{\mathbf{I}}_o + M_{vp}\dot{I}_p &= 0 \label{eq:vv}\\
        L_p\dot{I}_p + R_pI_p + M_{pa}\dot{\mathbf{I}}_a + M_{po}\dot{\mathbf{I}}_o + M_{pv}\dot{\mathbf{I}}_v &= 0 \label{eq:pp}
    \end{align}
\end{subequations}
where $M_{ij}$ are mutual inductance matrices between various conductors and the plasma ($a$ for shaping coils, $o$ for ohmic coils, $v$ for vacuum vessel, and $p$ for plasma), $R_i$ are resistances, $L_p$ is the plasma inductance, $I_i$ represent currents, and $V_i$ are voltages.

To generate the voltage feedforwards $\mathbf{V}_a$, given a desired shaping coil time derivative $\dot{\mathbf{I}}_{a}$, plasma current time derivative $\dot{I}_p$, and real-time estimates of $[\mathbf{I}_a, \mathbf{I}_o, \mathbf{I}_v, \mathbf{I}_p]$, the unknowns $[\mathbf{V}_a, \mathbf{V}_o, \dot{\mathbf{\mathbf{I}}}_o, \dot{\mathbf{\mathbf{I}}}_v]$ must be solved for. 
It is worth noting that the resistive term $R_a \mathbf{I}_a$ in eq.~\eqref{eq:aa} is already compensated by the inner coil current control loop, hence it can be discarded from the feedforward reference generation.
One additional consideration is that TCV features two independent ohmic-drive circuits designed so that the amount of poloidal stray field is minimal when their current is equal. To impose this additional constraint, let {$\dot{l}_o\in\mathbb{R}$ be a scalar and define}:
\begin{align}
    \mathbf{1}_2 \equiv \begin{bmatrix}
        1\\
        1
    \end{bmatrix} \quad \dot{\mathbf{I}}_o \equiv \dot{l}_o\mathbf{1}_2
\end{align}

Now observe that Equations \ref{eq:vv} and \ref{eq:pp} involve only two unknowns, namely $\dot{l}_o$ and $\dot{\mathbf{I}}_v$, which can be solved for with elementary algebra. Plugging these quantities into \ref{eq:aa} and \ref{eq:oo} yields the feed-forward voltages. The system of equations is given by:
\begin{subequations}
    \begin{align}
        c_1 &\equiv (-M_{po} + M_{pv}M_{vv}^{-1}M_{vo}) \mathbf{1}_2\\
        c_2 &\equiv M_{pv}M_{vv}^{-1}(R_v \mathbf{I}_v + M_{va}\dot{\mathbf{I}}_a + M_{vp}\dot{I}_p)\nonumber\\
        \dot{l}_o &= \frac{1}{c_1}(L_p\dot{I}_p + R_pI_p + M_{pa}\dot{\mathbf{I}}_a - c_2)\\
        \dot{\mathbf{I}}_o &\equiv \dot{l}_o\mathbf{1}_2\\
        \dot{\mathbf{I}}_v &= -M_{vv}^{-1}(M_{vo}\dot{\mathbf{I}}_o + R_v \mathbf{I}_v + M_{va}\dot{\mathbf{I}}_a + M_{vp}\dot{I}_p)\\
        \mathbf{V}_a &= M_{ao}\dot{\mathbf{I}}_o  + M_{av}\dot{\mathbf{I}}_v + M_{aa}\dot{\mathbf{I}}_a  + M_{ap} \dot{I}_p \\
        \mathbf{V}_o &= M_{oo}\dot{\mathbf{I}}_o + M_{ov}\dot{\mathbf{I}}_v + R_o \mathbf{I}_o + M_{oa}\dot{\mathbf{I}}_a + M_{op} \dot{I}_p
    \end{align}
\end{subequations}
Initial experiments resulted in radial instabilities that were correlated with the vacuum vessel model. While the issue was never properly root-caused, neglecting the vessel model from these equations resulted in the successful experiments in this work.

It is noteworthy that most of the inductance and resistance matrices are, for practical purposes, static, and thus can be computed pre-pulse. The key challenge involves the mutual inductances related to the plasma, $p$, which vary with the plasma current distribution. It is found that a simple linear regression of the form~\eqref{eq:regression} with respect to the plasma magnetic axis position $(r_A, z_A)$ can yield accurate predictions across a broad range of shapes, yielding $R^2 > 0.98$ for all coils.
\begin{align}\label{eq:regression}
    M_{pa} \approx A_{pa}\begin{bmatrix}
        r_A\\
        z_A
    \end{bmatrix} + b_{pa}\quad
    M_{po} \approx A_{po}\begin{bmatrix}
        r_A\\
        z_A
    \end{bmatrix} + b_{po} \,,
\end{align}
where the $A$ matrices and $b$ vectors are regression parameters. Thus, in real-time the mutual inductance matrices can be computed using the estimated magnetic axis position. Additional variable quantities include the plasma inductance $L_p$ and resistance $R_p$. In principle, real-time estimates of $L_p$ and $R_p$ can be used, but constant, typical values are employed in this work.

\subsection{Integration with 10kHz control}\label{subsec:10khz_control}
Coil current, vertical stabilization, vertical position $zI_p$, and plasma current $I_p$ are handled in the TCV experiments presented in this article through a digital 10kHz controller. 
An emulator of the legacy controller used for these tasks at TCV, known as ``hybrid''~\cite{lister1997control,mele2025design}, was implemented in a recently commissioned digital magnetic control framework, called the Fast MAGnetic controller (FMAG), allowing for flexible specification of linear time-varying (LTV) controllers. 
One major change relative to the legacy hybrid controller was the introduction of a coarse finite-element based observer for the plasma vertical position, which provided a significant improvement in the accuracy of the estimated plasma $z$ position compared to the legacy observer. Moreover, it was also found that radial position estimates obtained with the finite-element based observer could disagree with those coming from equilibrium reconstruction by several centimeters, especially during current ramps. This observation motivated the inclusion of the radial position controller into a slower 1kHz control loop which has access to real-time equilibrium reconstruction. Note that this is not possible with the vertical position instead, as this quantity is directly related to plasma vertical stabilization and thus requires a larger bandwidth.

\subsection{Controller Handover}\label{subsec:handover}
\begin{figure}[!htbp]
    \centering
    \includegraphics[width=0.7\linewidth]{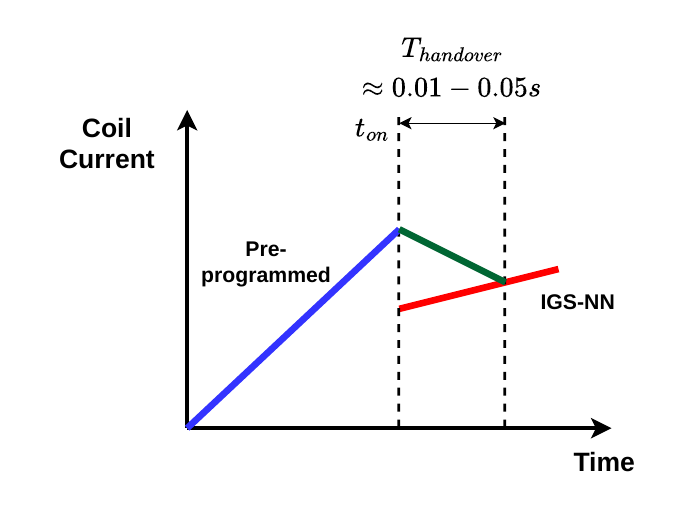}
    \caption{Depiction of the control hand-over strategy from the pre-programmed references (blue) to references generated by $IGS_{nn}$ (red) via a linear interpolation (green) over a handover window of $T_{handover}$ starting at time $t_{on}$.}
    \label{fig:control_handover}
\end{figure}

Since plasma break-down and burn-through are managed by the standard TCV control strategy, a handover from the break-down and early ramp-up phase to the new controller is needed. The adopted strategy was to have FMAG follow a pre-programmed reference trajectory until the controller activation time, $t_{on}$. After $t_{on}$, a linear interpolation is performed to smoothly transition from the pre-programmed reference to the $IGS_{nn}$ based reference over a handover window of $T_{handover}$. Both variables were set as tunable parameters. For a sense of typical values, the values used in pulse \#87970 were $t_{on}=0.175$s and $T_{handover}=0.025$s. Fig.~\ref{fig:control_handover} depicts the handover strategy.
\clearpage

\section{Summary of nomenclature}\label{app:nomenclature}

\begin{table}[htbp]
\centering
\footnotesize
\caption{Summary of symbols and notation used throughout this work.}
\label{tab:symb}
\begin{tabular}{ll@{\hspace{2em}}ll}
\hline
Symbol & Description & Symbol & Description \\
\hline
$\psi$                & Poloidal flux & $\beta_p$ & Poloidal beta \\
$j_\phi$, $j_e$        & Toroidal current density (plasma, ext. cond.) & $q_A$ & Safety factor on axis \\
$p(\psi)$             & Plasma pressure profile & $rB_t$ & Toroidal field function \\
$T(\psi)$             & Poloidal current function ($T = rB_t$) & $o_{\mathrm{diag}}$ & Magnetic diagnostics for RZIP control \\
$\bm{\theta}$          & Vector of plasma parameters & $r_{\mathrm{ref}}, z_{\mathrm{ref}}$ & Reference radial/vertical position \\
$\mathbf{I}_e$                 & Total external conductor currents & $I_p^{\mathrm{ref}}$ & Reference plasma current target \\
$\mathbf{I}_a$                 & Active coil currents (shaping + Ohmic) & $IGS_{nn}$ & NN surrogate of inverse Grad--Shafranov solver \\
$\mathbf{I}_s$              & Shaping coil currents & $o_t$ & Synthetic diagnostic observation (RL) \\
$\mathbf{I}_o$              & Ohmic drive coil currents & $O(\mathbf{I}_e,\bm{\theta})$ & Observation function generating $o_t$ \\
$\mathbf{I}_v$                 & Passive (vessel) conductor currents & $B_{\mathrm{stray}}$  & Poloidal stray field from $\mathbf{I}_o$, $\mathbf{I}_v$ \\
$I_p$                 & Plasma current & $\dagger_n$ & Truncated pseudo-inverse ($n$ largest s.v.) \\
$\mathbf{V}_a$                 & Voltages applied to active coils & $B_s$ & Matrix: shaping coil currents $\to$ poloidal field \\
$M_{ee}$              & Mutual inductance matrix of conductors & $B_o$, $B_v$ & Matrices: Ohmic/vessel currents $\to$ poloidal field \\
$R_e$                 & Diagonal resistance matrix of conductors & $\pi$ & Control policy \\
$\dot{\Psi}_{ep}$     & Induced EMF from plasma--conductor coupling & $d(\cdot,\cdot)$ & Shape error distance metric \\
$F(\mathbf{I}_e, \bm{\theta})$  & Grad--Shafranov solution operator & $S(\psi)$ & Shape observer function \\
$s$                   & Vector of plasma shape descriptors & $s^{\mathrm{targ}}$ & Target shape descriptor vector \\
\hline
\end{tabular}
\end{table}

\end{document}